\documentclass[preprint,showpacs,amssymb,aps ,groupedaddress,superscriptaddress ]{revtex4}
\usepackage{graphicx,amsmath}
\usepackage{bm}
\usepackage{epstopdf}
\begin{document}

\title{Impact of cut-off frequency effect on resonance energy transfer and Casimir-Polder interaction}

\author{Nguyen Dung Chinh}
\affiliation{Institute of Fundamental and Applied Sciences, Duy Tan University, 06 Tran Nhat Duat St., District 1, Ho Chi Minh City 70000, Vietnam}
\affiliation{Faculty of Environmental and Natural Sciences, Duy Tan University, 03 Quang Trung St., Hai Chau, Da Nang 50000, Viet Nam}
\email[]{nguyendungchinh@dtu.edu.vn}

\author{Vinh N.T. Pham}
\affiliation{Department of Physics \& Postgraduate Studies Office, Ho Chi Minh City University of Education, Ho Chi Minh City, Vietnam}

\author{Nguyen Duy Vy}\email[Corresponding author: ]{nguyenduyvy@vlu.edu.vn}  \affiliation{Laboratory of Applied Physics, Science and Technology Advanced Institute, Van Lang University, Ho Chi Minh City, Vietnam}
\affiliation{Faculty of Applied Technology, School of Technology, Van Lang University, Ho Chi Minh City, Vietnam}

\begin{abstract}
Using the Green's function approach, we investigate the resonance energy transfer (RET) rate between two parallel, identical two-level atoms in the presence of three types of cylindrical system: a distributed Bragg reflector (DBR), a perfectly reflecting wall (PRW), and a two-layer silicon fiber. Our analysis, incorporating the cut-off frequency condition, reveals significant suppression of the RET rate for atoms positioned along the axis of the cylinder with the PRW. In contrast, for atoms located within the DBR, the RET rate is enhanced in the far zone. Additionally, we find that for atoms oriented radially are placed inside or near the surface of the silicon fiber, the RET rate is entirely inhibited. We also investigate the Casimir-Polder (CP) interaction between a cut-off-frequency DBR and an excited atom, discovering a fully attractive potential towards the surface for the atom within the waveguide.
\end{abstract}

\maketitle
\section{Introduction}

Resonance energy transfer (RET) between an excited atom and a ground state atom involves the exchange of photons from the donor to the acceptor, mediated by the quantum electromagnetic field \cite{Craig84, Andrews99}. In free space, the RET rate follows a $r^{-6}$ dependence in the near zone (also known as Förster transfer) \cite{T}, where the separation between the two atoms is much smaller than the electronic energy transfer wavelength, $\lambda_{\rm A}$ ($r \ll \lambda_{\rm A}$). In this regime, electrostatic interactions dominate. In the far zone, where the interatomic distance significantly exceeds the transition wavelength ($r \gg \lambda_{\rm A}$), the RET rate exhibits a $r^{-2}$ dependence \cite{Avery66, Andrews92}.

When material bodies are present, the surrounding electromagnetic field is altered, affecting the radiative processes of quantum emitters \cite{Purcell46}. Recent studies have focused on the impact of highly symmetric systems on RET, due to their potential for controlling energy transfer. For instance, within macroscopic quantum electrodynamics (QED), the RET rate in the presence of dispersive and absorbing material bodies of arbitrary shapes has been analyzed \cite{Ho02}. This theory has been applied to multi-layer planar systems, revealing significant enhancements or suppressions in the RET rate for atoms positioned near planar interfaces.

The resonance energy transfer (RET) rate has been extensively studied using various approaches. The standard mode expansion has been applied to systems involving microspheres \cite{Leung88} and planar microcavities \cite{Kobayashi95, Cho95}. Another effective method is the Green's function formalism, which has been utilized to examine RET between emitters near diverse structures such as two-layer dielectric cylinders \cite{Marocico09}, coated metallic cylinders \cite{Karanikolas14}, planar waveguides \cite{Marocico11}, and graphene monolayers \cite{Karanikolas15}.

In addition to RET, the Casimir-Polder (CP) interaction, which is mediated by virtual photons, plays a crucial role when an atom is near a material body \cite{Buhmann12}. This interaction can lead to novel trapping mechanisms \cite{Chang14, Fermani07, Sambale08}. A general formula for calculating the CP interaction between an atom and an arbitrarily shaped dispersive and absorbing body has been derived using perturbation theory \cite{Buhmann04}. Subsequent research has explored the CP interaction in various contexts, including chiral atoms near surfaces at finite temperatures \cite{Barcellona16}, excited atoms near fiber tips \cite{Chinh18}, two-level atoms in Bragg-reflector cylindrical structures \cite{Chinh19}, atoms in optical cavities \cite{Fang19}, atoms near optical nanofibers \cite{Kien19, Kien22}, and Rydberg atoms near multilayered graphene structures \cite{Wong23}.

RET is a fundamental mechanism with significant implications in physical, chemical, and biological processes \cite{Sahoo11, Wu94}. Theoretical advancements have elucidated the mechanisms behind RET \cite{Craig84, Andrews99, Avery66, Andrews92, Gomberoff66, Zong19}. Enhancements in RET efficiency have been observed in polymer solar cells \cite{Huang13} and through the incorporation of SiO$_2$ layers between silver plates \cite{Basko00}. RET has also been utilized to assess excitation energy transfer between nanowires \cite{Salam19} and to enhance energy absorption in photodynamic therapy for cancer treatment \cite{Ling17}. Additionally, RET from excited dye states to graphene has been explored for measuring nanometer-scale geometric structures \cite{Swathi08}. A novel platform for detecting biomolecules via fluorescence RET from quantum dots to graphene oxide has been proposed \cite{Dong10}. Two-dimensional nanomaterials are increasingly used as donors and acceptors in Förster RET-based sensing, with applications spanning biochemical analysis, environmental monitoring, and disease diagnosis \cite{Zhou20}.

In this paper, we investigate a cylindrical structure composed of multiple layers with periodically varying permittivities between high and low values. This variation is designed such that the optical path length within each layer is a quarter of the wavelength, creating a distributed Bragg reflector (DBR) cylindrical waveguide. 
In such a case, phenomena related to atoms would be interesting when the atoms are placed in the innermost layer or outside the waveguide, since in those positions, the atoms can "see" the entire structure of the waveguide.
The effects of the DBR on the spontaneous decay rate of an excited two-level atom, the Casimir-Polder (CP) interaction of an atom within the DBR, and the resonance energy transfer (RET) rate between two emitters inside or outside the DBR have been previously studied \cite{Chinh19, Phuoc21}.

A counterpart of the DBR is a waveguide with perfectly reflecting wall (PRW) 
due to a similar characteristic: all light with frequencies surrounding 
the central frequency in the band-gap area \cite{Chinh19} can be reflected at the surfaces of the waveguides.
The RET rate of two atoms placed along the axis of a PRW has been explored with respect to the cut-off frequency, where the atomic transition frequency is below the waveguide’s cut-off frequency \cite{Passante18}. Building on this, we consider the RET rate between two atoms being on the axes of the DBR and examine the regions within the waveguides (DBR and PRW) where the cut-off frequency effect is significant. Additionally, we analyze the RET rate for atoms positioned inside or outside a silicon cylinder and investigate the CP interaction of an excited atom within the waveguides under the influence of the cut-off frequency. 
The difference between this paper and \cite{Chinh19, Phuoc21} is, therefore, the consideration of 
the cut-off frequency effect.

The paper is organized as follows. In the next section, we derive explicit expressions for the Green's functions corresponding to different dipole moment orientations. Section III presents and discusses the numerical results. The final section summarizes the key findings.

\section{Basic formulas}

\begin{figure}[!t!]
\noindent
\includegraphics[width=0.5\linewidth]{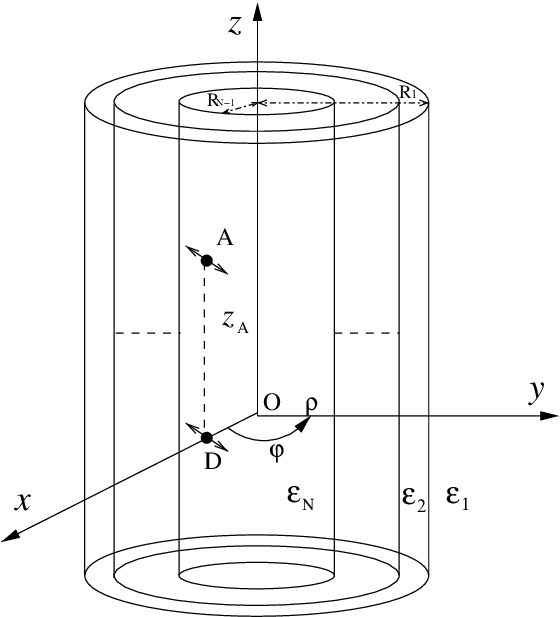}
\caption{
Two atoms in the presence of a multilayer cylinder.
}
\label{sys}
\end{figure}

Consider two two-level atoms, one in an excited state and the other in the ground state, 
surrounded by a multilayer dielectric cylinder with relative permeability equal to $1$ for simplicity 
(see Fig. \ref{sys}). 
Two atoms are assumed to be identical 
and in the same layer. 
They have transition frequency
$\omega_{\rm A}=\omega_{\rm D}$, electric dipole moment 
${\bf d}_{\rm A}={\bf d}_{\rm D}$,
positions 
${\bf r}_{\rm A}$=$(\rho_{\rm A}, \varphi_{\rm A}, z_{\rm A})$ and 
${\bf r}_{\rm D}$=$(\rho_{\rm D}, \varphi_{\rm D}, z_{\rm D})$. 
The resonance energy transfer (RET) rate from the donor atom to the acceptor one 
is determined by Green's tensor as \cite{Ho02}
\begin{equation}
\label{e1}
     \frac{\Gamma}{\Gamma_0} = 
     \biggl|
     \frac{
     {\bf d}_{\rm A} \cdot {\bm G}({\bf r}_{\rm A},{\bf r}_{\rm D},\omega) 
     \cdot {\bf d}_{\rm D}
     }
     {
     {\bf d}_{\rm A} \cdot {\bm G}^{(0)}({\bf r}_{\rm A},{\bf r}_{\rm D},\omega)
     \cdot {\bf d}_{\rm D}
     }
      \biggr|^2
      = \biggl| 
     \frac{ 
      \hat{\bf d}_{\rm A} \cdot {\bm G}^{(0)}({\bf r}_{\rm A},{\bf r}_{\rm D},\omega) 
      \cdot \hat{\bf d}_{\rm D} 
    + \hat{\bf d}_{\rm A} \cdot {\bm G}^{(fs)}({\bf r}_{\rm A},{\bf r}_{\rm D},\omega) 
    \cdot \hat{\bf d}_{\rm D}
     }
     {
     \hat{\bf d}_{\rm A} \cdot {\bm G}^{(0)}({\bf r}_{\rm A},{\bf r}_{\rm D},\omega) 
     \cdot \hat{\bf d}_{\rm D}
     }
      \biggr|^2      ,
\end{equation}
where $\Gamma_0$ is the RET rate in a homogeneous medium,
Green's function can be separated into two parts \cite{Li00}
\begin{equation}
\label{e2}
     {\bm G} ({\bf r},{\bf r}',\omega) = {\bm G}^{(0)} ({\bf r},{\bf r}',\omega) \delta^s_f
     + {\bm G}^{(fs)} ({\bf r},{\bf r}',\omega) ,
\end{equation}
\begin{align}
\label{G0}
     &\bm{G}^{(0)}({\bf r}_{\rm A},{\bf r}_{\rm D},\omega) = 
       - \frac{\delta({\bf u})}{3k^2} {\bm I}
       + \frac{k}{4\pi} (a{\bm I} - b\hat{\bf{u}}\otimes\hat{\bf{u}}) 
         e^{iq},
\\
\label{a(q)}
   &  a \equiv a(q) = \frac{1}{q} + \frac{i}{q^2} - \frac{1}{q^3},     
\\
\label{b(q)}
   &  b \equiv b(q) = \frac{1}{q} + \frac{3i}{q^2} - \frac{3}{q^3}  
\end{align}
with $k=\sqrt{\varepsilon(\omega)}\frac{\omega}{c}$, 
${\bf u}={\bf r}-{\bf r}'$, 
$f$ and $s$ are the layer indices of the coordinates of 
the field point ${\bf r}$ and the source point ${\bf r}'$,
$\hat{{\bf u}}={\bf u}/u$,
$q=ku$, $\delta_f^s$ is the Kronecker delta, and $\bm{I}$ being the unit tensor.

For ${\bf r}_{\rm A} = (\rho_{\rm A},0,z_{\rm A})$, 
${\bf r}_{\rm D} = (\rho_{\rm D},0,0)$.
The orientation of two atoms is set parallel as
\begin{align}
\label{e3}
    \hat{\bf d}_{\rm A} \equiv \hat{\bf z}, \hat{\bf d}_{\rm D} \equiv \hat{\bf z},
\\    
\label{e4}
    \hat{\bf d}_{\rm A} \equiv \hat{\bm \rho}_{\rm A}, 
    \hat{\bf d}_{\rm D} \equiv \hat{\bm \rho}_{\rm A},
\\
\label{e5}
    \hat{\bf d}_{\rm A} \equiv \hat{\bm \varphi}_{\rm A}, 
    \hat{\bf d}_{\rm D} \equiv \hat{\bm \varphi}_{\rm A}.
\end{align}
The respective components of the homogeneous Green's tensor read as
\begin{equation}
\label{e12}
     G_{zz}^{(0)}({\bf r}_{\rm A},{\bf r}_{\rm D},\omega_{\rm A}) = 
     \sqrt{\varepsilon(\omega_{\rm A})}\frac{k_{\rm A}}{4\pi}
     \left( a - b \frac{z^2_{\rm A}}{u^2} \right)
    e^{iq} ,
\end{equation}
\begin{align}
\label{e29}
      G_{\rho_{\rm A}\rho_{\rm A}}^{(0)}({\bf r}_{\rm A},{\bf r}_{\rm D},\omega_{\rm A})
     = \sqrt{\varepsilon(\omega_{\rm A})}\frac{k_{\rm A}}{4\pi}
     \left( a  
     - b \frac{(\rho_{\rm A} - \rho_{\rm D})^2
               }{u^2} \right)
        e^{iq} ,
\end{align}
\begin{equation}
\label{e30}
     G_{\varphi_{\rm A}\varphi_{\rm A}}^{(0)}({\bf r}_{\rm A},{\bf r}_{\rm D},\omega_{\rm A}) = 
     \sqrt{\varepsilon(\omega_{\rm A})}\frac{k_{\rm A}}{4\pi}
      a e^{iq} .
\end{equation}

The second term in Eq. (\ref{e2}) is the scattering Green's tensor,
for the source and the field positions in the last layer
$f=s=N$, which reads \cite{Li00}
\begin{align}
\label{G(NN)}
 {\bm G}^{(NN)} &({\bf r}_{\rm A},{\bf r}_{\rm D},\omega) =
     \frac{i}{8\pi} \int_{-\infty}^\infty {\rm d}h
     \sum_{e,o}\sum_{n=0}^\infty \frac{(2-\delta_n^0)}{\eta_N^2}
\nonumber\\
   & \times\left[C_{3H_o^e}^{NN} {\bf M}_{_o^e n\eta_N}(h) \otimes {\bf M}'_{_o^e n\eta_N}(-h)
    + C_{3V_o^e}^{NN} {\bf N}_{_o^e n\eta_N}(h) \otimes {\bf N}'_{_o^e n\eta_N}(-h) \right.
\nonumber\\
   & + \left. C_{4H_o^e}^{NN} {\bf N}_{_e^o n\eta_N}(h) \otimes {\bf M}'_{_o^e n\eta_N}(-h)
    + C_{4V_o^e}^{NN} {\bf M}_{_e^o n\eta_N}(h) \otimes {\bf N}'_{_o^e n\eta_N}(-h) \right],
\end{align}
where the vector wave function ${\bm M}$ and ${\bm N}$ are written in the 
cylindrical coordinate system as:
\begin{align}
\label{M}
     {\bm M}_{_o^e n\eta_f}(h) &= \left[\mp \frac{nJ_n(\eta_f\rho)}{\rho}
     {}_{\cos}^{\sin}(n\varphi) \hat{\bm \rho} - \frac{\partial J_n(\eta_f\rho)}{\partial\rho}
     {}_{\sin}^{\cos}(n\varphi) \hat{\bm \varphi} \right] e^{ihz} ,
\end{align}
\begin{align}
\label{N}
     {\bm N}_{_o^e n\eta_f}(h) &= \frac{1}{\sqrt{h^2+\eta_f^2}}
     \left[ih\frac{\partial J_n(\eta_f\rho)}{\partial \rho} {}_{\sin}^{\cos}(n\varphi) \hat{\bm \rho}
     \mp \frac{ihn}{\rho} J_n(\eta_f\rho) {}_{\cos}^{\sin}(n\varphi) \hat{\bm \varphi} \right.
\nonumber\\     
    & + \left.\eta_f^2 J_n(\eta_f\rho) {}_{\sin}^{\cos}(n\varphi) \hat{\bf z}
        \right] e^{ihz}  .
\end{align}
${\bm M}'$, ${\bm N}'$ denote the functions of the donor coordinate ${\bf r}_{\rm D}$ while 
${\bm M}$, ${\bm N}$ are functions of the acceptor coordinate ${\bf r}_{\rm A}$.
The coefficients $C$ are calculated according to \cite{Li00}:
\begin{equation}
\label{C(3,4)}
   \begin{bmatrix}
   C_{3(H,V)_o^e}^{NN}\\
   C_{4(H,V)_o^e}^{NN}
   \end{bmatrix}
   =
   \begin{bmatrix}
   T_{31}^{(H,V)_o^e} & T_{32}^{(H,V)_o^e}\\
   T_{41}^{(H,V)_o^e} & T_{42}^{(H,V)_o^e}
   \end{bmatrix}
   \begin{bmatrix}
   T_{11}^{(H,V)_o^e} & T_{12}^{(H,V)_o^e}\\
   T_{21}^{(H,V)_o^e} & T_{22}^{(H,V)_o^e}
   \end{bmatrix}^{-1}
   \begin{bmatrix}
   1\\
   0
   \end{bmatrix},
\end{equation}
\begin{equation}
\label{Tmatr}
     {\bm T}^{(H,V)_o^e} = 
     \left[{\bm F}_{N(N-1)}^{(H,V)_o^e}\right]^{-1}
       {\bm P}^{(H,V)_o^e}
     {\bm F}_{11}^{(H,V)_o^e},
\end{equation}
\begin{equation}
\label{Pmatr}
     {\bm P}^{(H,V)_o^e} = 
     {\bm P}_{N-1}^{(H,V)_o^e}
     {\bm P}_{N-2}^{(H,V)_o^e} 
     \cdots
     {\bm P}_{2}^{(H,V)_o^e},
\end{equation}
\begin{equation}
\label{Pmatr_l}
     {\bm P}_l^{(H,V)_o^e} = 
     {\bm F}_{ll}^{(H,V)_o^e}
     \left[{\bm F}_{l(l-1)}^{(H,V)_o^e}\right]^{-1},
\end{equation}
\begin{equation}
\label{FHmatr}
     {\bm F}_{jm}^{H^e_o} =
     \begin{bmatrix}
       \eta_j H'_{jm}
    &  \mp \frac{\eta_j\zeta_j}{n} \tilde{H}_{jm}
    &  \eta_j J'_{jm}
    &  \mp \frac{\eta_j\zeta_j}{n} \tilde{J}_{jm}   \\
       0
    &  \varrho_j \frac{\xi_{jm}}{n} \tilde{H}_{jm}
    &  0
    &  \varrho_j \frac{\xi_{jm}}{n} \tilde{J}_{jm}   \\
       \pm \frac{\eta_j\zeta_j \tau_j}{n} \tilde{H}_{jm}
    &  \tau_j\eta_j H'_{jm}
    &  \pm \frac{\eta_j\zeta_j \tau_j}{n} \tilde{J}_{jm}
    &  \tau_j\eta_j J'_{jm}  \\
       \tau_j\varrho_j \frac{\xi_{jm}}{n} \tilde{H}_{jm}
    &  0
    &  \tau_j\varrho_j \frac{\xi_{jm}}{n} \tilde{J}_{jm}
    &  0
     \end{bmatrix},
\end{equation}
\begin{equation}
\label{FVmatr}
      {\bm F}_{jm}^{V^e_o} =
      \begin{bmatrix}
       \pm \frac{\eta_j\zeta_j}{n} \tilde{H}_{jm}
    &  \eta_j H'_{jm}
    &  \pm \frac{\eta_j\zeta_j}{n} \tilde{J}_{jm}
    &  \eta_j J'_{jm}  \\    
       \varrho_j \frac{\xi_{jm}}{n} \tilde{H}_{jm}
    &  0
    &  \varrho_j \frac{\xi_{jm}}{n} \tilde{J}_{jm}
    &  0  \\      
       \tau_j\eta_j H'_{jm}
    &  \mp \frac{\eta_j\zeta_j \tau_j}{n} \tilde{H}_{jm}
    &  \tau_j\eta_j J'_{jm}
    &  \mp \frac{\eta_j\zeta_j \tau_j}{n} \tilde{J}_{jm}  \\   
       0
    &  \tau_j\varrho_j \frac{\xi_{jm}}{n} \tilde{H}_{jm}
    &  0
    &  \tau_j\varrho_j \frac{\xi_{jm}}{n} \tilde{J}_{jm}   
      \end{bmatrix},
\end{equation}
where 
$k_f = \sqrt{\varepsilon_f (\omega)} \frac{\omega}{c}$, 
$\eta_f = \sqrt{k_f^2 - h^2}$,
$\tau_j=\sqrt{\varepsilon_j(\omega)}$, 
$\zeta_j=ihn/k_j$, 
$\varrho_j=\eta_j^2/k_j$, 
$\xi_{jm} \equiv \eta_jR_m$,
$H_{jm} \equiv H_n^{(1)}(\xi_{jm})$
and
$J_{jm} \equiv J_n(\xi_{jm})$
are the Hankel and the Bessel function of the first kind, respectively,
$H'_{jm} \equiv \frac{dH_n^{(1)}(\xi_{jm})}{d\xi_{jm}}$, 
$J'_{jm} \equiv \frac{dJ_n(\xi_{jm})}{d\xi_{jm}}$,
$\tilde{H}_{jm} \equiv \frac{nH_n^{(1)}(\xi_{jm})}{\xi_{jm}}$,
$\tilde{J}_{jm} \equiv \frac{nJ_n(\xi_{jm})}{\xi_{jm}}$.
$H$ ($V$) represents transverse electric (magnetic) waves 
and $e$, $o$ are implied even and odd functions.
The components of the scattering Green's functions respective to the orientations 
(\ref{e3}), (\ref{e4}) read as
\begin{align}
\label{e38}
   G_{zz}^{(NN)}&({\bf r}_{\rm A},{\bf r}_{\rm D},\omega) = 
   \frac{i}{4\pi} \int_0^\infty {\rm d}h
     \sum_{n=0}^\infty (2-\delta_n^0) \cos(hz_{\rm A})
     C_{3V_{e}}^{NN} \frac{\eta_{\rm N}^2}{k_N^2}
     J_{n}(\eta_{\rm N}\rho_{\rm A}) J_{n}(\eta_{\rm N}\rho_{\rm D}),
\end{align}
\begin{align}
\label{e39}
  G_{\rho_{\rm A}\rho_{\rm A}}^{(NN)}&({\bf r}_{\rm A},{\bf r}_{\rm D},\omega) =
     \frac{i}{4\pi} \int_{0}^\infty {\rm d}h
     \sum_{n=0}^\infty (2-\delta_n^0) \cos(h z_{\rm A})
\nonumber\\
   & \times\left[C_{3H_e}^{NN}  
                \tilde{J}_{NA}\tilde{J}_{ND}   
               + C_{3V_e}^{NN} 
                 \frac{h^2}{k_N^2}   
                 J'_{NA}J'_{ND} \right.
\nonumber\\
   & \left. - C_{4H_e}^{NN} 
       \frac{ih}{k_N}   
       J'_{NA}\tilde{J}_{ND} 
     - C_{4V_e}^{NN} 
        \frac{ih}{k_N}  
        \tilde{J}_{NA} J'_{ND}  \right] ,
\end{align}
\begin{align}
\label{e40}
 G_{\varphi_{\rm A}\varphi_{\rm A}}^{(NN)}&({\bf r}_{\rm A},{\bf r}_{\rm D},\omega)=
     \frac{i}{4\pi} \int_{0}^\infty {\rm d}h
     \sum_{n=0}^\infty (2-\delta_n^0) \cos(h z_{\rm A})
\nonumber\\
   & \times\left[C_{3H_e}^{NN} J'_{NA} J'_{ND}   
    + C_{3V_e}^{NN} \frac{h^2}{k^2_N} 
                     \tilde{J}_{NA} \tilde{J}_{ND} \right.
\nonumber\\
   & \left. - C_{4H_e}^{NN} \frac{ih}{k_N} \tilde{J}_{NA} J'_{ND}  
    - C_{4V_e}^{NN} \frac{ih}{k_N} J'_{NA} \tilde{J}_{ND} \right] ,
\end{align}
where
$J'_{NA} \equiv \frac{dJ_n(\eta_N\rho_A)}{d(\eta_N\rho_A)}$,
$\tilde{J}_{NA} \equiv \frac{nJ_n(\eta_N\rho_A)}{\eta_N\rho_A}$,
$J'_{ND} \equiv \frac{dJ_n(\eta_N\rho_D)}{d(\eta_N\rho_D)}$,
$\tilde{J}_{ND} \equiv \frac{nJ_n(\eta_N\rho_D)}{\eta_N\rho_D}$ .
Here we used the properties of $C_{3H_e}^{NN} = C_{3H_o}^{NN}$, 
$C_{3V_e}^{NN} = C_{3V_o}^{NN}$,
$C_{4H_e}^{NN} = - C_{4H_o}^{NN}$,
$C_{4V_e}^{NN} = - C_{4V_o}^{NN}$.
When the atoms are outside the cylinder, i.e., in layer 1,
equations (\ref{e38})-(\ref{e40}) remain valid after replacing 
$G^{(NN)} \to G^{(11)}$, $C_3^{NN} \to C_1^{11}$, $C_4^{NN} \to C_2^{11}$,
$k_N \to k_1$, $\eta_N \to \eta_1$,
$J_{n}(\eta_{\rm N}\rho) \to$ the Hankel function of the first kind 
$H^{(1)}_{n}(\eta_1\rho)$.
The coefficients $C_1^{11}$, $C_2^{11}$ are given by
\begin{equation}
\label{C(1,2)}
   \begin{bmatrix}
   C_{1(H,V)_o^e}^{11}\\
   C_{2(H,V)_o^e}^{11}
   \end{bmatrix}
   =-
   \begin{bmatrix}
   T_{11}^{(H,V)_o^e} & T_{12}^{(H,V)_o^e}\\
   T_{21}^{(H,V)_o^e} & T_{22}^{(H,V)_o^e}
   \end{bmatrix}^{-1}
   \begin{bmatrix}
   T_{13}^{(H,V)_o^e}\\ 
   T_{23}^{(H,V)_o^e}
   
   \end{bmatrix}.
\end{equation}

When two atoms are on the axis of the cylinder (i.e. $\rho_{\rm A} = \rho_{\rm D} = 0$), 
the Bessel function takes an approximate form as \cite{Abramowitz}
\begin{equation}
\label{Jlim}
     J_n(\xi) \sim \frac{1}{n!} \left(\frac{1}{2}\xi\right)^n.
\end{equation}
By substituting this form into Eqs. (\ref{e38}), (\ref{e39}), and (\ref{e40}), 
only the ($n=0$)-terms survive in Eq. (\ref{e38}) and only the ($n=1$)-terms 
survive in Eqs. (\ref{e39}) and (\ref{e40}), we end up with
\begin{align}
\label{e38_1}
   G_{zz}^{(NN)}&({\bf r}_{\rm A},{\bf r}_{\rm D},\omega) = 
   \frac{i}{4\pi} \int_0^\infty {\rm d}h
     \cos(hz_{\rm A})
     C_{3V_{e}}^{NN} \frac{\eta_{\rm N}^2}{k_N^2},
\end{align}
\begin{align}
\label{e39_1}
  G_{\rho_{\rm A}\rho_{\rm A}}^{(NN)}({\bf r}_{\rm A},{\bf r}_{\rm D},\omega) &=
  G_{\varphi_{\rm A}\varphi_{\rm A}}^{(NN)}({\bf r}_{\rm A},{\bf r}_{\rm D},\omega) = 
\nonumber\\  
   &  \frac{i}{8\pi} \int_{0}^\infty {\rm d}h
       \cos(h z_{\rm A})
   \left[C_{3H_e}^{NN}  
               + C_{3V_e}^{NN} 
                 \frac{h^2}{k_N^2}  
                 - 2 C_{4H_e}^{NN} 
       \frac{ih}{k_N}   
        \right] .
\end{align}
Here we used the properties of 
$C_{4H_e}^{NN} = C_{4V_e}^{NN}$.

For a cylindrical system with a perfectly reflecting wall,
one can treat the cylinder as two-layer system $N=2$ 
with $\varepsilon_2=1$, $\varepsilon_1 \to -\infty$,
$\eta_1^2 \to k_1^2$. In that case, we obtain 
$C_{3H_e}^{NN} = -\frac{H'_{21}}{J'_{21}}$,
$C_{3V_e}^{NN} = -\frac{H_{21}}{J_{21}}$,
$C_{4H_e}^{NN} = C_{4V_e}^{NN} = 0$ \cite{Haakh15,Ho16}.

\section{Numerical results and discussion}

\subsection{Resonace energy transfer rate}

\begin{figure}[!t!] \centering
\includegraphics[width=0.7\linewidth]{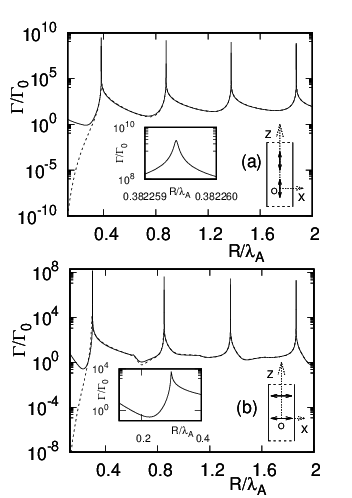}
\caption{
The normalized RET rate, relative to the free-space value, as a function of cylindrical radii for atoms located along the axes of the waveguides. The distance between the two atoms is $z = 1\lambda_{\rm A}$. Panels (a) and (b) show results for atoms with dipoles oriented parallel to and orthogonal to the waveguide axis, respectively. Solid lines correspond to the distributed Bragg reflector (DBR) with $N = 15$, $\varepsilon_H = 9$, $\varepsilon_L = 1$, while dashed lines represent results for the perfectly reflecting wall (PRW). As the radii of the waveguides increase, several distinct poles appear. The difference between the DBR and PRW results is noticeable only in a very small range of radii.
}
\label{fig2}
\end{figure}

In Fig. \ref{fig2}, we present the normalized resonance energy transfer (RET) rate between two atoms positioned along the axis of two cylindrical systems. The normalization is relative to the RET rate in a vacuum, where one atom is excited and the other is in the ground state.

The first system is a distributed Bragg reflector (DBR) comprising 15 layers, with permittivities periodically varying between a high value of $\varepsilon_H = 9$ and a low value of $\varepsilon_L = 1$. Each layer’s optical path length is set to a quarter wavelength (i.e. $R_j - R_{j+1} = \lambda_{\rm A}/[4{\rm Re}\sqrt{\varepsilon_{j+1}}]$) 
\cite{Ho03, Tran11}, 
with the innermost layer being vacuum. This system configuration was selected based on \cite{Chinh19}, where the significant contrast in permittivities and the number of layers are sufficient to produce a band gap (see Fig. 2 in \cite{Chinh19}). Although the permittivities here are frequency-independent, the results for a silicon-silica cylindrical system in \cite{Chinh19} confirm the reasonableness of our choice.

The second system is a cylindrical waveguide with a perfectly reflecting wall (PRW). We set the distance between the two atoms to $1\lambda_{\rm A}$. The results for the DBR are depicted by solid lines, while the PRW results are shown by dashed lines.

In panel (a), where the dipoles are aligned parallel to the waveguide axes (axial dipoles), the RET rate displays resonant poles, with the number of poles increasing as the cylinder radius grows. 
This trend is consistent with the results for spontaneous decay rates reported in \cite{Chinh19}.
Comparing the DBR with the PRW, we observe that the TM modes contribute to the poles labeled $p_{01}$, $p_{02}$, $p_{03}$, and $p_{04}$ from left to right \cite{Chinh19}.

In panel (b), where the dipoles are oriented orthogonally to the waveguide axes, four poles are identified as $q_{11}$, $q_{12}$, $q_{13}$, and $q_{14}$, corresponding to TE modes. 
For atoms positioned off the axis, additional poles such as $p_{11}$ and $p_{12}$ (not shown) appear with increasing radius. Despite similar pole positions for both the DBR and PRW systems, the peak values for the PRW cannot be obtained due to the divergence of the Green’s functions at these pole positions \cite{Chinh19}.
On the other hand, the peaks for the DBR can be obtained similarly to the first one, 
as shown in the inset of Fig. 2(a), which follows a Gaussian distribution.

It is evident that these sharp peaks result from choosing real permittivities. For comparison, we replaced these real permittivities with a complex one, $\varepsilon_H = 9 + 0.1i$, $\varepsilon_L = 1$, introducing an absorptive property to the medium. As shown in the inset of Fig. \ref{fig2}(b), the peak is more softened compared to the peaks caused by real permittivities.

In this paper, we focus on the cut-off frequency effect, which arises when the atomic transition frequency is lower than the waveguide's cut-off frequency
\begin{equation}
\label{40}
\omega_{\rm A} < \omega_{min}.
\end{equation}
Here $\omega_{min}$ is defined 
as \cite{Passante18}
\begin{align}
\label{e41}
        (\omega_{min})_{TM} \simeq \frac{2.4c}{R} ,
\\
\label{e42}
         (\omega_{min})_{TE} \simeq \frac{1.8c}{R} .
\end{align}
These frequencies correspond to the first two poles ($p_{01}$, $q_{11}$) shown
in Fig. \ref{fig2} (a), (b).
The condition (\ref{40}) results in the regime $k_{\rm A}R < 1.8$ or $R/\lambda_{\rm A} < 0.28$.
In the presence of the cut-off frequency effect, the RET rates for two waveguides, 
the DBR and the PRW, show up differences 
in the region 
$0.125 \leq R/\lambda_{\rm A}<0.28$ (see Fig. \ref{fig2}(b)).

\begin{figure}[!t!]
\noindent
\includegraphics[width=0.7\linewidth]{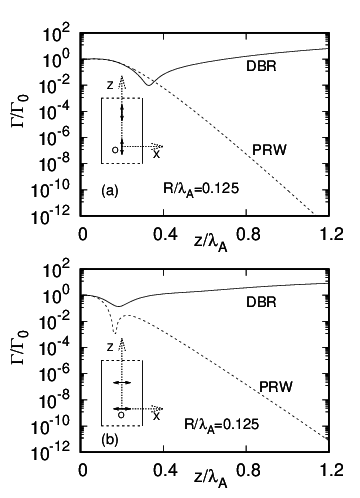}
\caption{
The normalized RET rate as a function of the distance between two atoms positioned along the axes of the waveguides. The curves are shown for the distributed Bragg reflector (DBR) with $N = 15$, $\varepsilon_H = 9$, 
$\varepsilon_L = 1$ (solid line) and for the perfectly reflecting wall (PRW) (dashed line). Panels (a) and (b) show results for dipoles oriented parallel to and orthogonal to the axis, respectively. The radii of the waveguides are $R_{14} = R = 0.125\lambda_{\rm A}$.
The curves diverge by up to twelve orders of magnitude due to the influence of the cut-off frequency.
}
\label{fig3}
\end{figure}

We now examine the rate of resonance energy transfer (RET) between two dipole moments, normalized relative to the free-space value (Fig. \ref{fig3}). The two atoms are positioned along the axes of the waveguides. The distributed Bragg reflector (DBR), shown by solid lines, has the same characteristics as in Fig. \ref{fig2}, while the perfectly reflecting wall (PRW) is depicted by dashed lines.

The normalized RET rate is plotted as a function of the interatomic distance, considering both the near zone ($z<\lambda_{\rm A}$) and in the far zone ($z>\lambda_{\rm A}$).
When the dipoles are very close to each other, the curves approach unity, as expected. 
This is because, in general, the interaction between particles or bodies 
tends to approach their free space values, whether perturbing objects are taken far away 
or the particles are brought close together.

In our study, the innermost radius of the DBR and the PRW is set to $0.125\lambda_{\rm A}$ ($R_{14} = R = 0.125\lambda_{\rm A}$), meaning that the optical path length of each layer is a quarter of the atomic transition wavelength. In this regime, the cut-off frequency effect becomes significant.

Figures \ref{fig3}(a) and (b) show the results for dipoles oriented parallel and orthogonal to the waveguide axes, respectively. In the very near zone, Fig. \ref{fig3}(a) for $0<z/\lambda_{\rm A} \lesssim 0.2$ and Fig. \ref{fig3}(b) for $0<z/\lambda_{\rm A} \lesssim 0.1$, the results for both systems are in good agreement. The cut-off frequency effect is notably significant in both the near and far zones, as evidenced by the distinct curves.

The RET rate between two atoms within the PRW waveguide shows a suppression, 
while for the DBR, there is an enhancement. 
Both systems exhibit inhibition in the very near zone, with the suppression being more pronounced for orthogonal dipoles in the PRW compared to parallel dipoles. These findings are consistent with the energy transfer amplitude results reported in \cite{Passante18}. Additionally, we have reproduced Fig. 3 from \cite{Passante18}, which studied the amplitude of RET between two axial dipoles in the PRW waveguide with $R/\lambda_{\rm A}=0.02$ in
the near zone.

\begin{figure}[!t!] \centering
\includegraphics[width=0.96\linewidth]{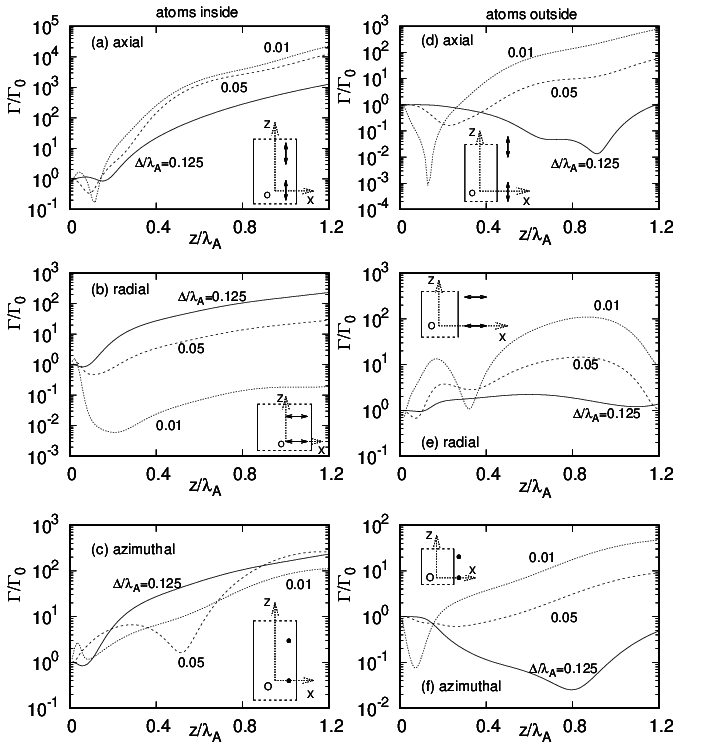}
\caption{
The normalized RET rate as a function of interatomic distance for different dipole moment orientations and atom-surface distances $\Delta$.
The cylindrical system consists of two layers, with permittivities $\varepsilon_1 = 1$, $\varepsilon_2 = 12.25$ (representing silicon). Atoms are positioned in the $Oxz$-plane along a line parallel to the cylinder axis, either inside the waveguide (left panels (a), (b), and (c)) or outside the waveguide (right panels (d), (e), and (f)). The waveguide radius is $R/\lambda_{\rm A} = 0.125$. 
RET is completely suppressed as dipoles approach the surface, but this suppression occurs only for radial dipole moments inside the fiber.
}
\label{fig4}
\end{figure}

In Fig. \ref{fig4}, we examine the normalized RET rate for two atoms positioned either inside (left-side panels) or outside (right-side panels) a two-layer cylinder with permittivities $\varepsilon_1 = 1$, $\varepsilon_2 = 12.25$, representing a silicon optical fiber. The RET rate is plotted as a function of interatomic separation, with the atoms aligned along a line parallel to the fiber axis. The fiber radius is $R/\lambda_{\rm A} = 0.125$, matching the innermost radius of the DBR and the radius of the PRW considered earlier.

We analyze three atom-surface distances: $\Delta/\lambda_{\rm A} = 0.125$ (solid lines), $0.05$ (dashed lines), and $0.01$ (dotted lines). 
When the atoms are very close to each other, the RET rate approaches unity, reflecting dominant atomic interaction over interaction with the fiber wall.

Inside the fiber:

$\bullet $
Axial dipoles (Fig. \ref{fig4}(a)): For atoms on the axis ($\Delta/\lambda_{\rm A} = 0.125$, solid line), the RET rate slightly decreases, reaching a minimum in the near zone
($z/\lambda_{\rm A} \simeq 0.2$) before increasing sharply by roughly three orders of magnitude in the far zone ($z/\lambda_{\rm A} = 1.2$). This effect is more pronounced for atoms off the axis. For $\Delta/\lambda_{\rm A} = 0.05$ and $0.01$, the minima 
shift to $z/\lambda_{\rm A} \simeq 0.15$ and become deeper compared to the on-axis case, with RET rates rising by more than four orders of magnitude in the far zone.

$\bullet $
Radial and azimuthal dipoles (Figs. \ref{fig4}(b) and (c)): The results for radial and azimuthal dipoles show identical solid line curves. These curves experience slight suppression in the very near zone ($z/\lambda_{\rm A} \simeq 0.1$)
but exhibit a steady increase in the RET rate for the rest of the range ($0.1 \lesssim z/\lambda_{\rm A} \leq 1.2$). Unlike the axial dipole case, the RET rate for radial dipoles near the surface ($\Delta/\lambda_{\rm A} = 0.01$), is completely inhibited. The distinction arises because axial and azimuthal dipoles are parallel to the surface in the near-boundary limit, while radial dipoles are perpendicular.

Outside the fiber:

$\bullet $
General behavior (Figs. \ref{fig4}(d), (e), (f)): For atoms located outside the silicon cylinder, when far from the surface ($\Delta/\lambda_{\rm A} = 0.125$),
the RET rates oscillate around unity for all dipole orientations, as shown by the solid lines. Compared to the symmetric case where atoms are aligned with the fiber axis, the impact of the fiber on the RET rate is less significant when atoms are outside the fiber. For atoms close to the surface, the RET rate is inhibited in the near zone but enhanced in the far zone for axial and azimuthal dipoles. In contrast, for radial dipoles at $\Delta/\lambda_{\rm A} = 0.01$,  the RET rate is uniformly enhanced in both the near and far zones.

\subsection{Resonant Casimir-Polder potential}

\begin{figure}[!t!] \centering
\includegraphics[width=1.\linewidth]{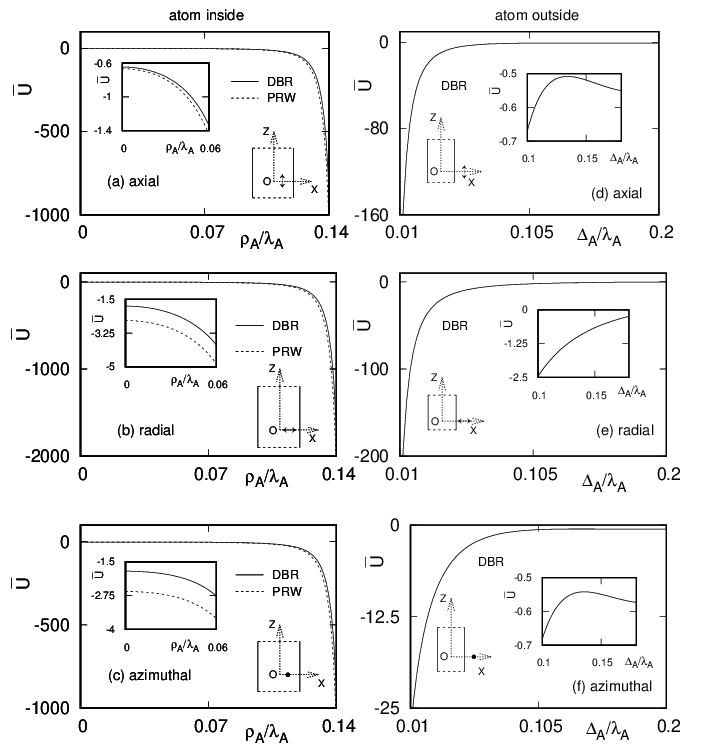}
\caption{
The dimensionless CP potential $\bar{U} = U8\pi c/[\mu_0\omega_{\rm A}^3 d_{\rm A}^2]$ as a function of the atomic radial position (left panels, atom inside the waveguides) and the atom-surface distance (right panels, atom outside the DBR) for different dipole moment orientations. The cylindrical systems have radii $R_{14} = R = 0.15\lambda_{\rm A}$, 
with other parameters of the DBR matching those in Fig. \ref{fig3}. Oscillations observed at the distance $\Delta_{\rm A}/\lambda_{\rm A} = 0.15$ confirm the effect of the cut-off frequency.
}
\label{fig5}
\end{figure}

For a two-level excited atom in the presence of an arbitrary material body, the Casimir-Polder (CP) interaction exerted by the body on the atom includes a resonant component, which is described by \cite{Buhmann04}
\begin{equation}
\label{U}
     U=-\mu_0\omega_{\rm A}^2 d_{\rm A}^2
     {\rm Re}\ \hat{\bf d}_{\rm A}{\bm G}^{(fs)}({\bf r}_{\rm A},
     {\bf r}_{\rm A},\omega_{\rm A}) \hat{\bf d}_{\rm A}.
\end{equation}
%
Here, we only need the scattering part of the Green's function at the atom's position.
When the excited atom is located in the innermost layer of a multi-layer cylinder, 
we replace the Green's tensors in equations (\ref{e38})–(\ref{e40}) 
in equation (\ref{U}) by substituting ${\bf r}_{\rm D} \to {\bf r}_{\rm A}$ and setting $z_{\rm A} \to 0$.

In Figs. \ref{fig5}(a), (b), and (c), we investigate the Casimir-Polder (CP) interaction potentials between an excited atom and a DBR (solid lines) or a PRW (dashed lines) waveguide as the atom is moved from the center towards the surfaces of the waveguides. The potentials are plotted as a function of the atomic radial position.  
The DBR have 15 layers ($\varepsilon_H = 9$, 
$\varepsilon_L = \varepsilon_{15} = 1$).
The radius of the innermost layer of the DBR is equal to the radius of the PRW 
$R_{14} = R = 0.15 \lambda_{\rm A}$,
where the cut-off frequency effect is present.

Overall, as the atom is moved from the center to a position 
separated by $0.01 \lambda_{\rm A}$ 
from the surface, it experiences an attractive interaction, as shown in Figs. \ref{fig5}(a), (b), and (c). The insets in the left-side panels reveal that the potentials are fully attractive. Among the different orientations, the radial orientation results in the deepest potential compared to the axial and azimuthal orientations.

In the presence of the cut-off frequency effect, the interactions of the excited atom with both the DBR and the PRW are quite similar. The surfaces tend to pull the atom towards them, leading to an adsorption phenomenon. This is consistent with the findings for waveguide radii $R_{14} = R = 0.125 \lambda_{\rm A}$,
discussed in Fig. \ref{fig3}. This behavior contrasts with the results in \cite{Chinh19}, where a dip in the CP potential at the center of the waveguide was observed due to the $P_{01}$  mode, suggesting that the atom is pushed towards the cylinder axis.

The CP interaction between an excited atom and the surface of a macroscopic body can be either attractive or repulsive, depending on the atom-surface distance and the surrounding medium. For instance, \cite{Chinh18} found that the CP potential is repulsive (attractive) for an atom inside (outside) a two-layer dielectric cylinder ($\varepsilon_1 = 1$, $\varepsilon_2 = 1.1$) when it is close to the surface. For atoms farther from the surface, the CP potential oscillates between attraction and repulsion.

To determine whether the attractive potentials observed in Figs. \ref{fig5}(a), (b), and (c) are due to the close-surface effect or the cut-off frequency effect, we examine the atom's interaction when
it is placed outside the DBR. Figs. \ref{fig5}(d), (e), and (f) show the CP potential as a function of the atom-surface distance  $\Delta_{\rm A}$.
At $\Delta_{\rm A}/\lambda_{\rm A} = 0.15$, which corresponds to the atom's central position inside the DBR, the CP potentials shift from attractive to repulsive for axial and azimuthal dipole moments (see the insets in Figs. \ref{fig5}(d) and (f)). This transition indicates that the monotonic attractive potentials observed inside the waveguide are primarily due to the cut-off frequency effect.

This result can be explained by the fact that the waveguides cannot support the smallest mode ($R/\lambda_{\rm A} \simeq 0.29$, Fig. \ref{fig2}(b)) at the cut-off frequency. In contrast, all wavelengths resonant with $\lambda_{\rm A}$ exist outside the waveguides and can interfere at considerable distances.

\section{Summary}
In this paper, we investigated the resonance energy transfer (RET) rate between an excited atom and a ground-state atom with parallel dipole moments. Our focus was on atoms positioned in the innermost layer of a multi-layer cylindrical waveguide, specifically a distributed Bragg reflector (DBR). We also analyzed the behavior of atoms within a cylindrical waveguide featuring a perfectly reflecting wall (PRW), which serves as a counterpart to the DBR.

We observed that increasing the radius of the waveguides while positioning atoms along the axis leads to the appearance of resonant poles at specific radii, indicating a significant enhancement in the RET rate. This enhancement can reach magnitudes of approximately ten orders of magnitude for dipoles parallel to the DBR axis. These resonant positions align well for both DBR and PRW cases. However, in the small range $0 < R/\lambda_{\rm A} < 0.28$, the results diverge by several orders of magnitude between the two systems.

We considered the effect of the cut-off frequency, which notably differentiates the results for DBR and PRW by up to twelve orders of magnitude. When the waveguides are replaced by a two-layer silicon fiber, an enhancement in the RET rate is observed for atoms approaching the surface from both inside and outside the fiber, except for radial orientations inside the fiber, where the RET rate is completely suppressed.

Finally, the Casimir-Polder (CP) interaction between the DBR or PRW and an excited atom in the innermost layer, considering the cut-off frequency effect, reveals a fully attractive potential towards the surface of the waveguides.


\section*{Acknowledgments}

We are grateful to Ho Trung Dung for suggesting the problem and for expert guidance. 
This research is funded by Vietnam National Foundation for Science and Technology Development (NAFOSTED) under grant number 103.01-2023.36.



\end{document}